# Searching for states analogous to the $^{12}$C Hoyle state in heavier nuclei using the thick target inverse kinematics technique

M. Barbui[a], K. Hagel[a], J. Gauthier[a], S. Wuenschel[a], R. Wada[a], V.Z. Goldberg[a], R. T. deSouza[b], S. Hudan[b], D. Fang[c], X-G Cao[c,a] and J.B. Natowitz[a]

[a] *Cyclotron Institute, Texas A&M University, MS3366 College Station, TX*
[b] *Indiana University, Bloomington, IN, USA*
[c] *Shanghai Institute of Applied Physics (SINAP), Chinese Academy of Sciences, Shanghai, China*

**Abstract:** Identification of alpha cluster states analogous to the $^{12}$C Hoyle state in heavier alpha-conjugate nuclei can provide tests of the existence of alpha condensates in nuclei. Such states are predicted for $^{16}$O, $^{20}$Ne, $^{24}$Mg, $^{28}$Si etc. at excitation energies slightly above the multi-alpha particle decay threshold, but have not yet been experimentally identified. The Thick Target Inverse Kinematics (TTIK) technique can be used to study the breakup of excited self-conjugate nuclei into many alpha particles. The reaction $^{20}$Ne+α was studied using a $^{20}$Ne beam at 12 MeV/nucleon from the K150 cyclotron at Texas A&M University. The TTIK method was used to study both single α-particle emission and multiple α-particle decays. Events with alpha multiplicity up to four were analyzed. The analysis of the three α - particle emission data allowed the identification of the Hoyle state and other $^{12}$C excited states decaying into three alpha particles. The results are shown and compared with other data available in the literature. Although the statistics for events with alpha multiplicity four is low, the data show a structure at about 15.2 MeV that could indicate the existence in $^{16}$O of a state analogous to the $^{12}$C Hoyle state. This structure is confirmed by the re-analysis of alpha multiplicity four events from a previous experiment performed at 9.7 MeV/nucleon with a similar setup but lower granularity. Moreover, the reconstructed excitation energy of $^{24}$Mg for these events peaks at around 34 MeV, very close to the predicted excitation energy for an excited state analogous to the $^{12}$C Hoyle state in $^{24}$Mg.

**Introduction:**

The alpha cluster structure of nuclei with an equal number of protons and neutrons (alpha conjugate nuclei) was categorized in 1968 by Ikeda [1] to explain some excited states not reproduced by the shell model. Since then many studies have been performed, but alpha clustering is still not completely understood especially in the medium-light and heavy systems. In the last ten years a lot of theoretical effort has been focused on the study of $0^+$ states built on alpha-particle cores in self-conjugate nuclei at excitation energies slightly above the multi-alpha particle decay threshold. Those states are described as diluted gases of alpha particles occupying the same $0^+$ orbital. They are characterized by a larger radius compared to the normal bound states so that the interaction between alpha particles is reduced. Therefore, these states can be considered as the best candidates for Bose-Einstein condensates of alpha particles in the atomic nucleus. Examples of such states are the ground state of $^{8}$Be and the famous $^{12}$C Hoyle state. Analogous states are predicted for $^{16}$O, $^{20}$Ne, $^{24}$Mg, $^{28}$Si etc. at excitation energies slightly above the multi-alpha particle decay threshold, but have not yet been experimentally identified [2, 3].

Yamada et al. [4] estimated a maximum limit for such states of 10 alpha particles ($^{40}$Ca), resulting from the competition of the short range attractive force between alpha particles and the Coulomb repulsion as the number of alpha particles and the radius of the system increase.Proving the existence of alpha cluster states analogous to the $^{12}$C Hoyle state in heavier alpha-conjugate nuclei can provide a way to prove the existence of alpha condensates in nuclear matter. Kokalova et al. suggest that the signature for multi-alpha condensed states would be the decay of the excited system into pieces that are themselves condensates, ie. $^8$Be$_{gs}$, $^{12}$C$_{Hoyle}$, etc.[5]. Great experimental progress has recently been made on understanding the structure of the $^{12}$C Hoyle state [6-11] through the observation of the $2^+$ state in the Hoyle-state rotational band [6, 8, 9]. Some experimental work has been performed on heavier systems. Funaki et al. [12] predicted a $0^+$state in $^{16}$O at 15.1 MeV (the $0_6^+$ state) with a $^{12}$C "Hoyle" state structure coupled to an alpha particle. Excited states in$^{16}$O above the four alpha decay threshold have been studied by several authors [13-15]. Few studies exist on $^{20}$Ne [16], $^{24}$Mg [17, 18].

In this paper we investigate the existence of multi-alpha condensed states in self-conjugate nuclei heavier than $^{12}$C using the Thick Target Inverse Kinematics (TTIK) [19]. This technique is suited for this purpose because it allows the exploration of a large range of incident energies in the same experiment. In inverse kinematics, the reaction products are focused at forward angles and can be detected with detectors covering a relatively small portion of the solid angle in the forward direction. Since we stop the beam in the gas target volume we can detect the lighterparticles emitted at zero degrees. We studied the reaction $^{20}$Ne + α at a maximum energy of 12 MeV/nucleon. The TTIK method was used to study both single α-particle emission and multiple α-particle decays. Events with alpha multiplicity up to four were analyzed. The results show that this technique can be successfully used to study the breakup of excited self-conjugate nuclei into many alpha particles.

**Experimental Setup:**

The experimental setup used in this experiment is shown in Figure 1. A 13 MeV/nucleon$^{20}$Ne beam was delivered by the K150 Cyclotron at Texas A&M. The beam entered the pressurized chamber through a 12.7 micron Havar® window. The pressure of the$^4$He gas in the chamber was adjusted in order to stop the beam approximately 5 cm before the detectors. The $^4$He gas acted as the target. Four 5x5 cm$^2$ΔE-E telescopes were placed at the end of the pressurized chamber at -12°, -4.7°, 3°, 13° degrees from the beam positionto detect the reaction products. The ΔE detectors consisted of 55 micron, double sided, 16x16 strip silicon detectors. Three of the E detectors were 1 mm thick quadrant silicon detectors; one was a 1 mm thick double sided, 16x16 strip silicon detector. The signals from the front strips of the ΔE detectors were sent to high gain Indiana University pre-amplifiersand digitized using Struck SIS1366 Flash ADCs. Those digitizers provided energy and time information relative to the cyclotron radio frequency. Particle identification was obtained from the two dimensional scatter plots of ΔE-E. The signals from the back strips of the double sided, 16x16 strip silicon ΔEs and from the E detectors, processed with high gain pre-amplifiers, and shaping amplifiers were then acquired with peak sensing ADCs. The time difference between alpha particles was used to select correlated alpha particles. We selected multiple alpha events in which the time difference between detected alpha particles was less than 50 ns. A small surface barrier silicon detector was placed inside the grazing angle to serve as a beam monitor detector. We determined that a total of 3.82 10$^{10}$ beam particles ($^{20}$Ne) were delivered to the experiment during this run.

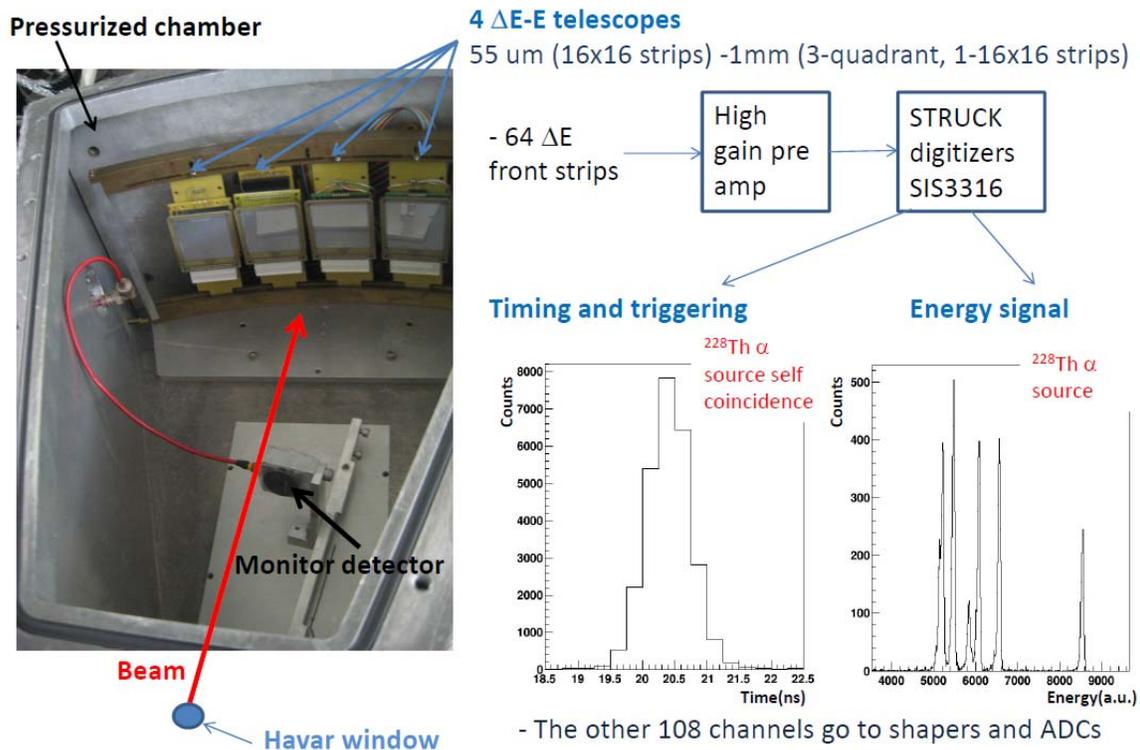

*Figure 1: Experimental setup and scheme of the electronics. Laboratory tests performed using a $^{228}$Th alpha source show that the signal processing with the STRUCK digitizers SIS3316 provides good energy and time resolution.*

**Alpha multiplicity one events:**

Events with alpha multiplicity one were selected to look for alpha resonant states in $^{24}$Mg. Due to the relatively high energy of the beam the density of states in the compound nucleus is quite high. Therefore, to be detectable, resonances must emerge from a continuum spectrum resulting from the statistical alpha evaporation from the excited $^{24}$Mg. In order to estimate the contribution of this continuum spectrum we have used PACE4 [20] and HIPSE [21] simulations. The calculations were performed varying the energy of the beam in steps of 0.5 MeV/nucleon, starting from 12 down to 4 MeV/nucleon. Our previous data [22] showed that at the energy of 3 MeV/nucleon the spectra are dominated by the resonant elastic scattering. In order to compare the simulated data with the experiment, we corrected the calculated alpha particle energy for the energy loss in the residual thickness of $^4$He gas depending on the estimated interaction point and we filtered the result with the detector geometry. Figure 2 shows the experimental alpha particle spectra per telescope compared with the calculated results. The simulated spectra are normalized in order to match the high energy tail of the experimental distributions. The experimental energy threshold of about 8 MeV is not applied to the simulation. Alpha particles with energy larger than 50 MeV punch through the entire telescope thickness and their total energies were reconstructedusing the SRIM code [23]. It is clear from Figure 2 that the experimental spectra show clear structures only at very small angles (Telescope 3) and these structures fade away as the detector angle increases. The alpha energy spectra measured in Telescopes 1 and 4 are completely determined by statistical evaporation. The PACE4 fusion-evaporation model is not able to reproduce the low energy part of the spectra recorded in Telescopes 2 and 3, suggesting that these alpha-particles are produced by other inelastic processes. The HIPSE code, which includes collision dynamics, seems to better reproduce the low energy part of the spectrum. In the following analysis we used the HIPSE calculation for Telescope 3 as best estimate of the continuum background.

After subtracting the continuum spectrum the data from the most central strips of Telescope 3 were analyzed to extract the excitation energy of the $^{24}$Mg in case of resonant elastic scattering. A drawback of using a thick target is that the position of the interaction point inside the gas volume is not directly measured and consequently we must determine that point, the energy of the beam at that point and the angle of the emitted particle. To achieve this, a reconstruction code was developed specifically for this experiment. The code is based on the procedure described in detail in Ref. [24]. Under the assumption that the production mechanism is resonant elastic scattering, the position of interaction point inside the gas volume is calculated in a recursive way, using the reaction kinematics, the measured alpha particle energies and range energy tables from SRIM. The code also provides the sum of the times of flight of the beam from the entrance window to the interaction point and of the alpha particle from the interaction point to the detector. This time is used to set the time zero for the measured time of flight and to correct for an observed time slewing with the amplitude of the signal.

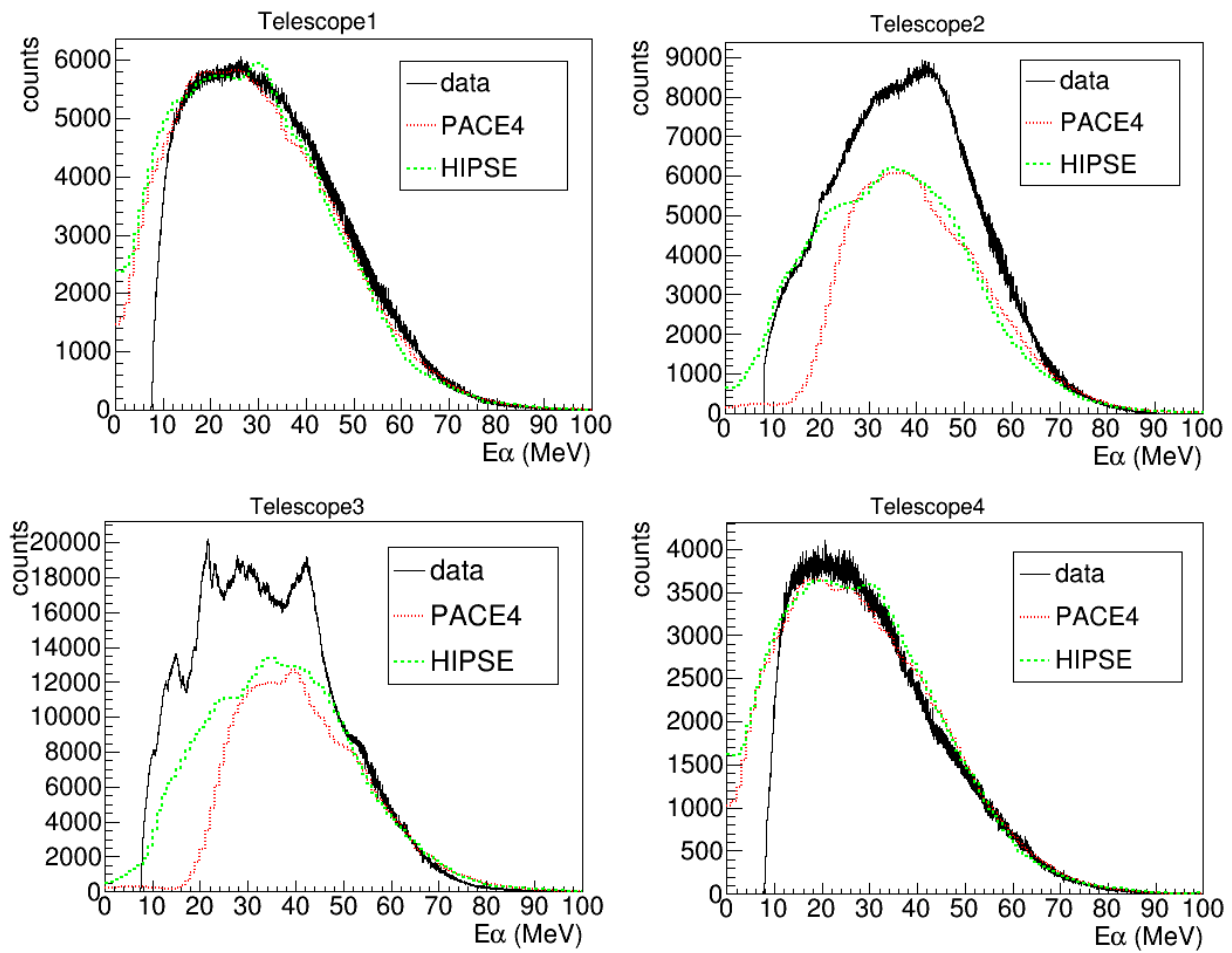

*Figure 2: Alpha particle energy spectra measured per telescope. PACE4 and HIPSE calculations are also shown.*

After reconstructing the position of the interaction point in the gas we obtained the differential cross-sections for resonant elastic scattering, in the center of mass frame as a function of the $^{24}$Mg excitation energy. The reconstruction is limited to the particles that did not punch trough the telescope. The cross section measured in Strip 2 of Telescope 3 corresponding to an average reconstructed angle of 5º is plotted in Figure 3 and compared with the data measured by Abegg and Davis [25], in normal kinematics at 168º in the center of mass using an alpha particle beam and a Ne target. Their measurement was carried out in energy steps of 10-15 keV. Even though the presentenergy resolution is worse than that of Ref [25], our technique allows obtaining the gross features of the spectrum in a single run. The Excitation energy spectrum is also extended up to 24 MeV in the present work. Many new resonances are observed.

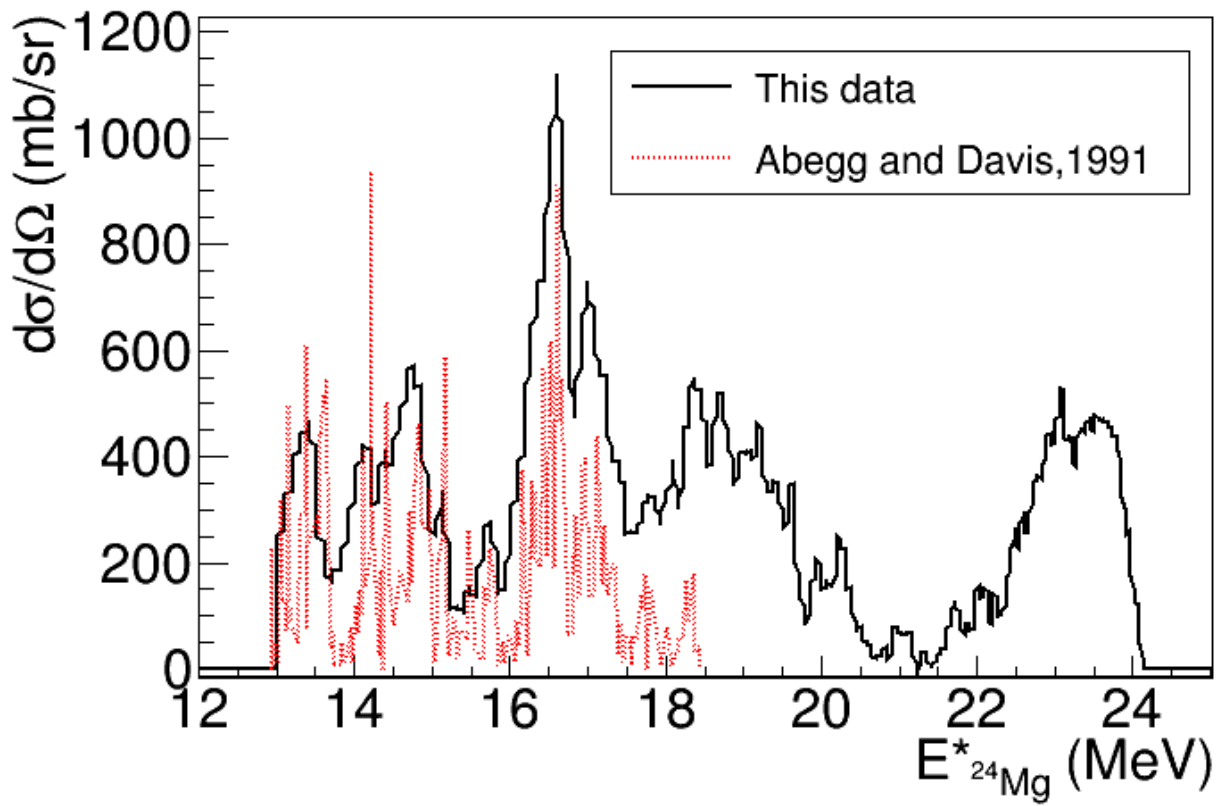

*Figure 3: Differential cross section for resonant scattering in the center of mass frame. The black line shows the result of this work, the red line is from Ref. [25]*

**Alpha multiplicity two events:**
Alpha multiplicity two events detected in a coincidence window of 50 ns were selected for the analysis. The focus of this section is on alpha multiplicity two events corresponding to the decay of $^8$Be in the ground or excited state. In order to highlight the possible correlation between the two detected alpha particles, we plot their energies against each other in Figure 4. The contribution due to statistical evaporation of two alpha particles, estimated using the HIPSE code, is also presented in Figure 4. The HIPSE result in Figure 4 is normalized to the data in order to have the same intensity in the area delimited by 70<E1<100 MeV and 10<E2<100 MeV. The bottom panel on Figure 4 shows the bin by bin difference when the HIPSE estimate of the background is subtracted from the data.

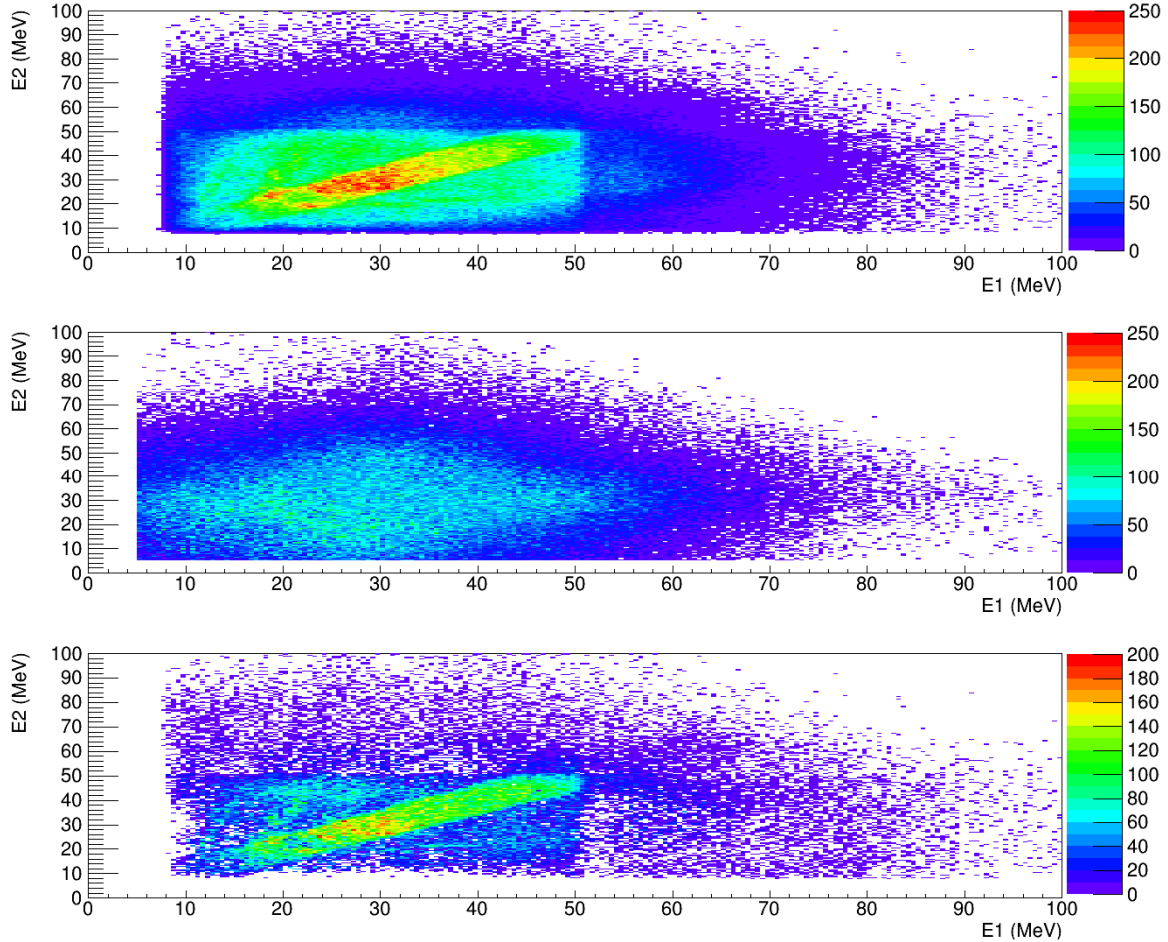

*Figure 4: Correlation between kinetic energies of the two detected alpha particles. The top panel shows the experimental data, the middle panel shows results of the HIPSE calculation after correcting for the energy loss in the gas the bottom panel shows the remainder when the events in the middle panel are subtracted from the data.*

The most important difference between the experimental data and the results of the HIPSE simulation is the band around the bisector. We will show in the following that these events comes from the decay of the $^8$Be ground state. The position of the interaction point inside the gas volume is reconstructed using a recursive procedure based on the reaction kinematics and energy and momentum conservation. The assumption is that $^{20}$Ne and $^4$He interact and give origin to $^8$Be and $^{16}$O. We assume that the $^{16}$O is always in the ground state while the $^8$Be can be excited. On each iteration, the kinetic and excitation energies of the $^8$Be are obtained from the kinetic energies of the two alpha particles after energy loss correction. At the end of the process, the quality of the reconstruction is double-checked by comparing the time of flight calculated during the reconstruction with the measured time of flight.

Figure 5 shows distribution of the experimental relative energy between two measured alpha particles. The background due to statistical evaporation estimated with HIPSE and PACE4 is shown. These backgrounds are obtained after passing the original HIPSE and PACE4 energies through our reconstruction code. Another background estimate is obtained by randomly mixing alpha particles from different multiplicity two events. The three background estimates are qualitatively similar. The events at relative energy greater than 5 MeV can be explained by the statistical evaporation, while the peaks at 92 keV and 2.9 MeV correspond to the energies of the ground state and first excited state in $^8$Be. It is important to note that the error on the determination of the $^8$Be excitation energy caused to an incorrect determination of the interaction point is minimal. In fact, the two detected alpha particles have similar and quite large energies in the laboratory system and travel similar fight-paths so that energy loss correction almost cancels out when we calculate the $^8$Be excitation energy. To prove this point we selected the events corresponding to the $^8$Be ground state by requiring the relative energy of the two alpha particles to be less than 200 keV; for those events we recalculated the relative energy assuming a fixed interaction point at 27 cm from the window. The result in Figure 6 shows that the determination of the relative energy is weakly influenced by the reconstructed position of the interaction point. Figure 7 shows the relative energy spectra obtained after subtracting the different background estimates. The ground state of $^8$Be is clearly visible and has the same magnitude no matter what background is used. The peak at 2.9 MeV emerges only after subtracting the HIPSE or PACE4 backgrounds and it is not visible after subtracting the mixed events background. This difference is probably explained by self-correlation in the mixed event data due to the large width of the 2.9 MeV state.

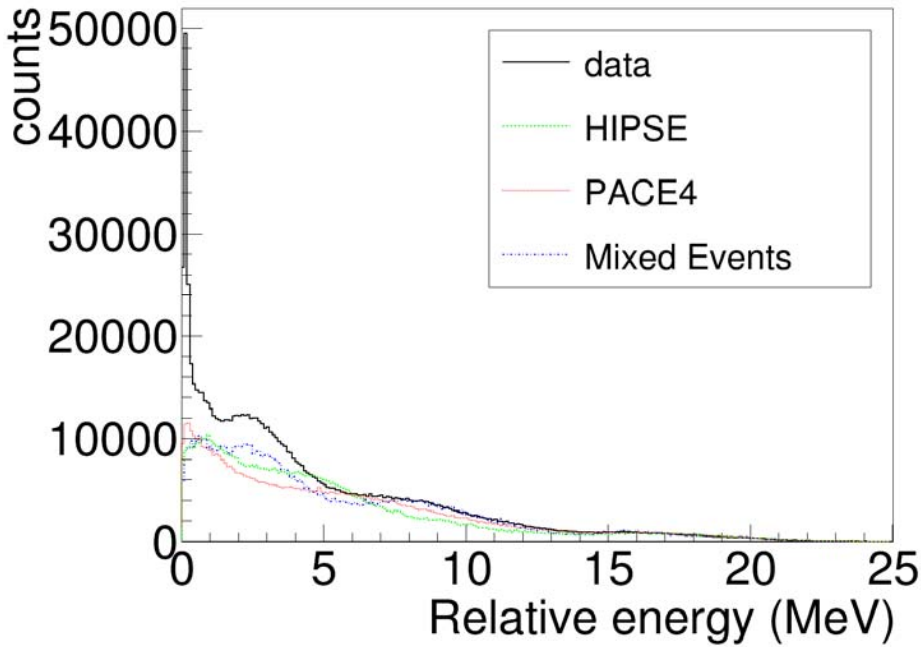

*Figure 5: Distribution of relative energy of two detected alpha particles. The background estimates obtained by HIPSE, PACE4 and by randomly mixing two alpha particles from different multiplicity two events are also shown.*

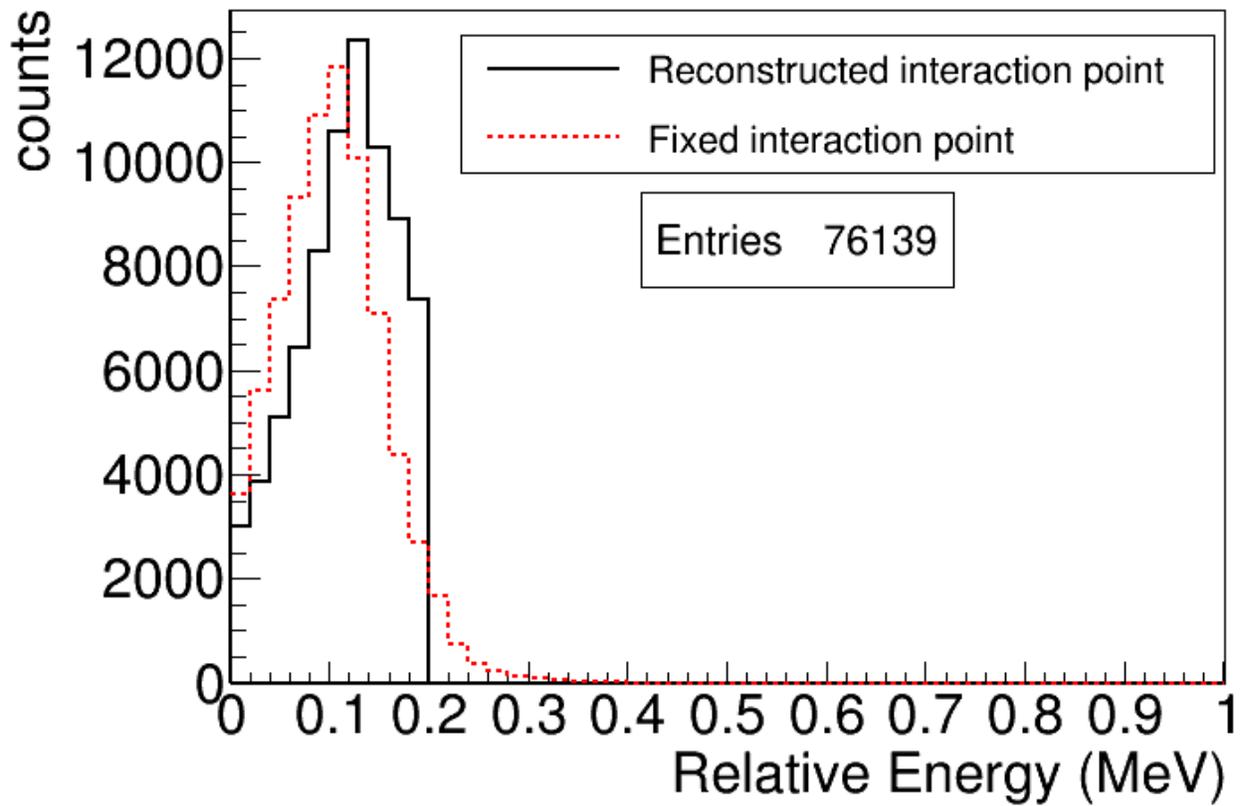

*Figure 6: Relative energy of two alpha particles. Black solid line: Interaction point obtained from the reconstruction code. Red dashed line: Fixed interaction point at 27 cm from the window, corresponding to beam energy of 6 AMeV.*

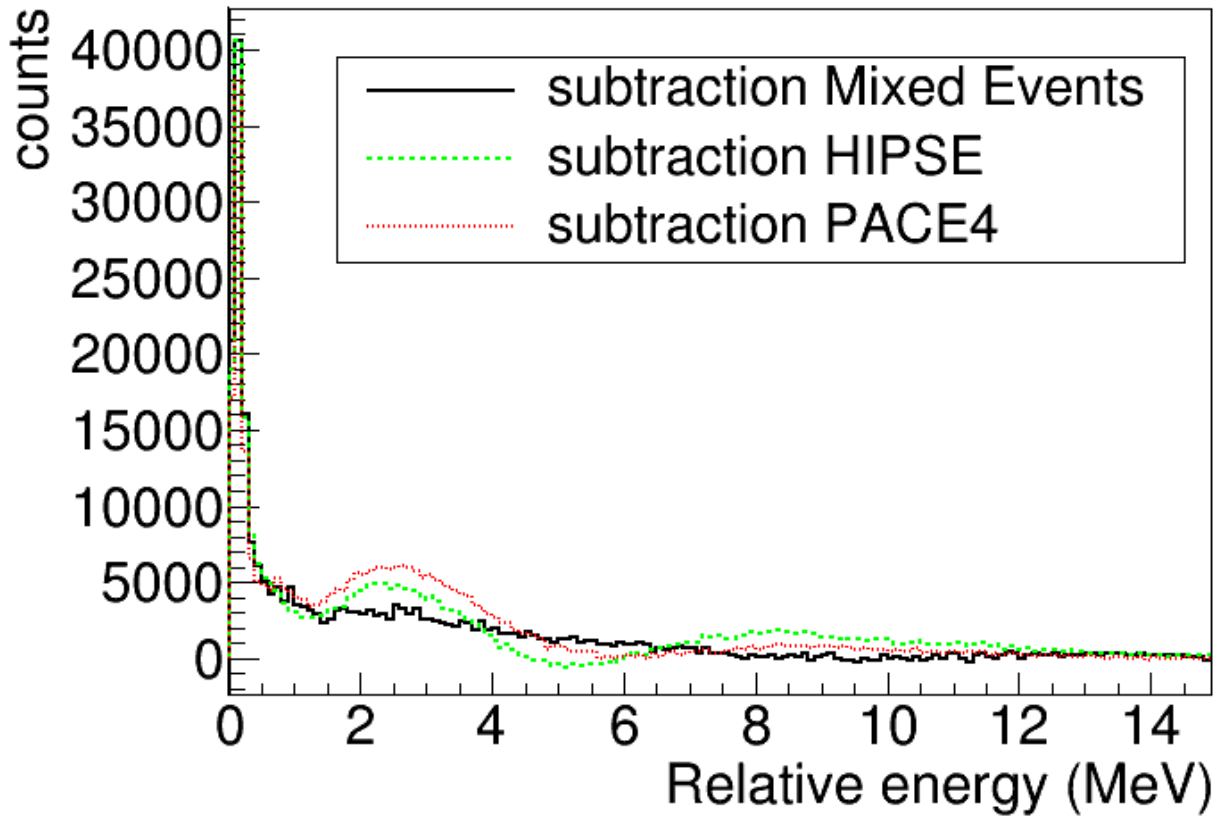

*Figure 7: Relative energy of two alpha particles after background subtraction.*

The left side of Figure 8 shows the two-dimensional plots E1 vs E2 obtained after selecting the events corresponding to the ground state of $^8$Be (top) and the ($2^+$) state (bottom). The experimental kinetic energy distributions for the same selections are shown in the right side of Figure 8. Some interesting structures are visible in the kinetic energy distribution of $^8$Be in the ground state. These might indicate that these $^8$Be are emitted in the decay of specific states in $^{24}$Mg, although other direct processes might also be responsible for these structures. A more detailed knowledge of the reaction process is needed to disentangle these two possibilities. No structures are visible in the kinetic energy spectra of $^8$Be in the $2^+$ state.

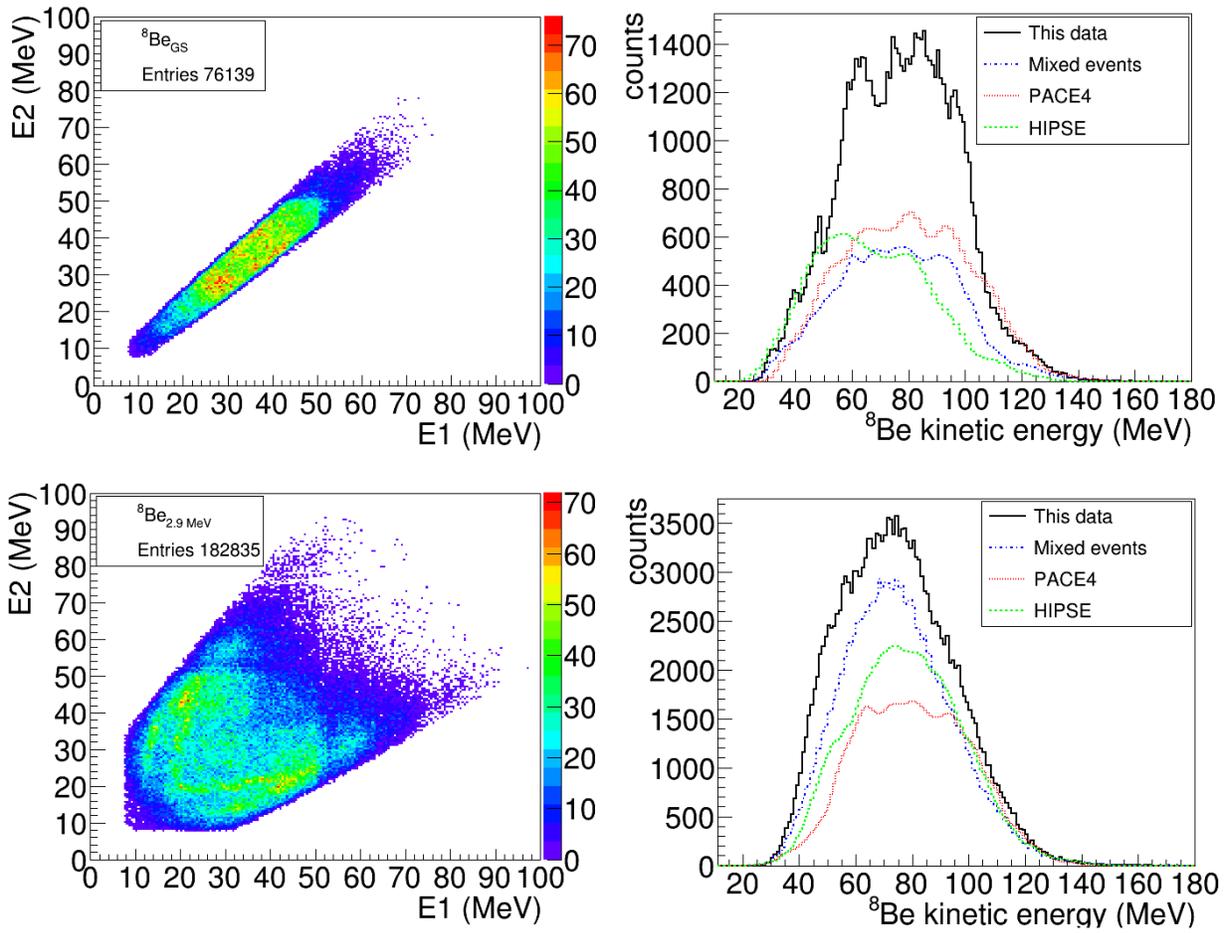

*Figure 8: Left panels: E1 vs. E2 plot obtained after selecting the events corresponding to the decay of the $^8$Be in the ground state (top) and in the $2^+$ state at 2.9 MeV (bottom). Right panels: kinetic energy of the $^8$Be in the ground state (top) and in the $2^+$ state at 2.9 MeV (bottom); background estimates from mixed events, PACE4 and HIPSE are also shown.*

**Alpha multiplicity three events:**

The analysis of the three α - particle emission data allowed the identification of the Hoyle state and other $^{12}$C excited states decaying into three alpha particles. Some preliminary results were reported in refs. [22, 26] and compared with data available in the literature. Events with 3 alpha particles arriving at the detectors within a time window of 50 ns were selected for the analysis. Since the detected alpha particles are correlated in time and position we can reasonably assume that in most cases they are coming from an excited $^{12}$C decaying into three alpha particles. Also in this case the position of the interaction point inside the gas volume is obtained using a recursive procedure based on the reaction kinematics and energy and momentum conservation. The assumption is that two $^{12}$C nuclei are produced in the interaction, one in the ground state, the other with enough excitation energy to split in three alpha particles. On each iteration, the kinetic energy of the $^{12}$C is obtained from the kinetic energy of the 3 alpha particles after energy loss correction; the excitation energy of the $^{12}$C splitting into three alpha particles is obtained from the sum of the kinetic energies of the three alpha particles in their center of mass plus the $Q_{gg}$ value. Also in this case, as for the multiplicity two events, the error on the determination of the $^{12}$C excitation energy due to an incorrect determination of the interaction point is minimal. The contribution due to the statistical evaporation of three alpha particles was estimated using HIPSE and PACE4. Figure 9 shows the $^{12}$C excitation energy calculated using the reconstructed interaction point. The background estimated with HIPSE, PACE4 and by randomly mixing alpha particles from different events is also shown. All the background estimates are normalized to the high energy tail of the data. The background estimated by the mixed events is much larger than the HIPSE or PACE4 estimated in the region from 9 to 15 MeV, this is probably due to the presence of a broad resonance around 10 MeV in $^{12}$C. In order to emphasize the narrow Hoyle and (3$^-$) states we decided to subtract the mixed event background in the following analysis.

Figure 10 shows the excitation energy spectrum of $^{12}$C obtained after subtracting the background resulting from the random mixing of alpha particles from different multiplicity 3 events. The Hoyle state and the (3$^-$) state at 9.64 MeV, clearly distinguishable in the spectrum, were analyzed in more detail. These states were selected by gating on the $^{12}$C excitation energy windows (7.36, 7.76) MeV for the Hoyle state and (9.57, 9.88) MeV for the (3$^-$) state. The relative energies of the three possible couples of alpha particles were calculated event by event, in order to determine if the decay proceeded through the ground state of $^8$Be or not. Figure 11 shows the spectra of the relative energy for the three possible couples of alpha particles, for the Hoyle state and the (3-) state. For these two states the smallest value of the relative energy peaks around 92 keV, showing that the decay proceeded through the $^8$Be ground state. The relative energies of the three possible couples of alpha particles resulting from a Monte Carlo simulation of the Hoyle state and the 3$^-$ state decaying through the $^8$Be ground state are reported in Figure 12 and agree very well with the experimental results. Figure 11 also shows the Dalitz plots for the Hoyle state and the 3- state. The corresponding Dalitz plots, obtained from a Monte Carlo simulation of the Hoyle state and the 3$^-$ state decaying through the $^8$Be ground state, are shown in Figure 12. The experimental Dalitz plots show the characteristic shape for the sequential decay through the $^8$Be ground state.

These results are compatible with the latest experimental results on the decay of the Hoyle state and the (3$^-$) state in $^{12}$C [27-29]

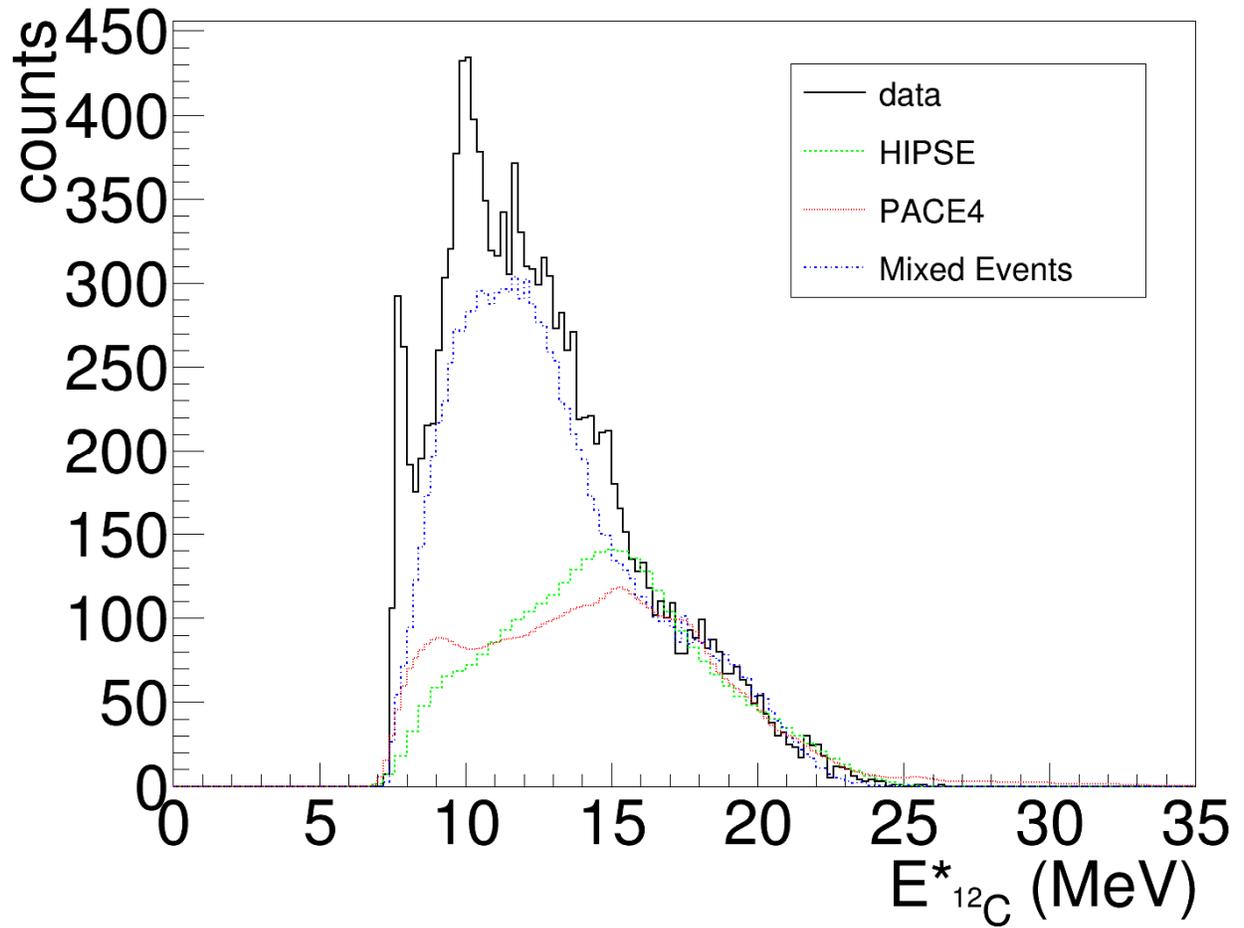

*Figure 9: Excitation energy of $^{12}C$ obtained from the alpha multiplicity 3 events. The backgrounds estimated by HIPSE, PACE4 and mixed events are also plotted.*

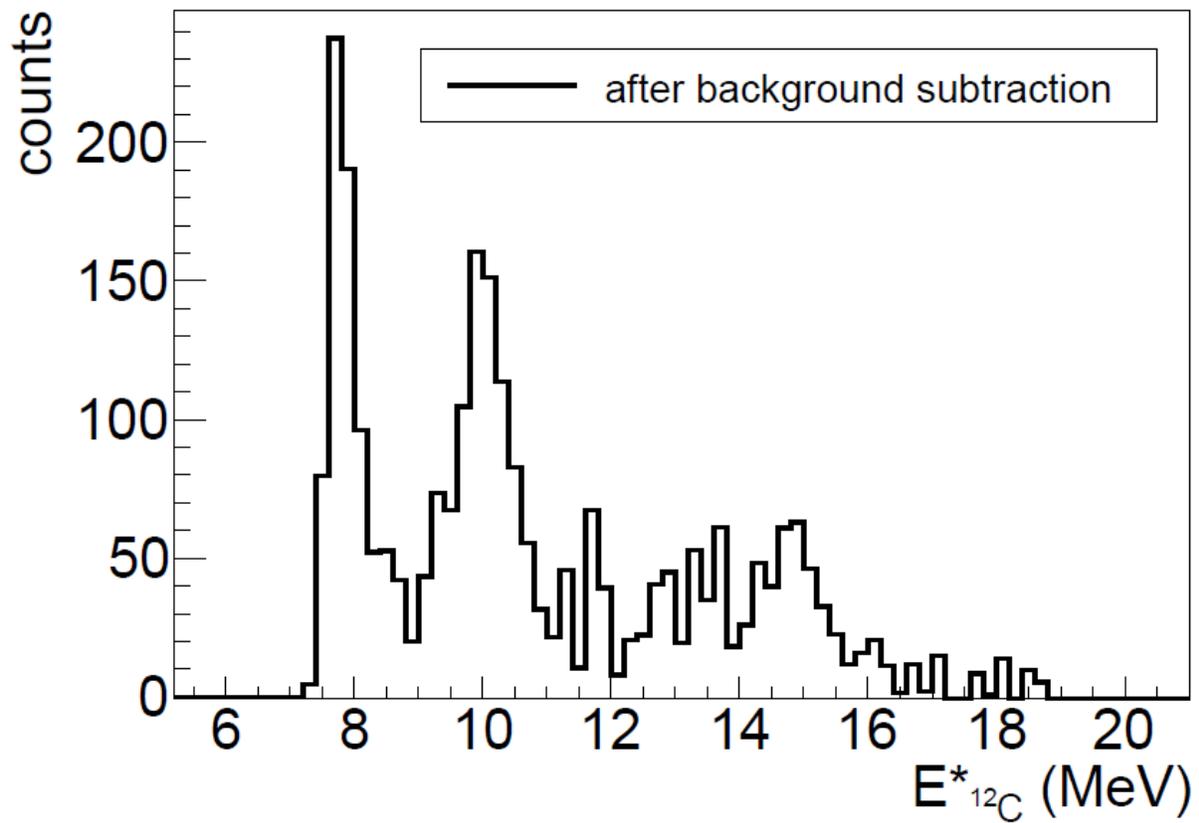

Figure 10: Excitation energy of $^{12}C$ obtained from the alpha multiplicity 3 events obtained after subtracting the background from the mixed events.

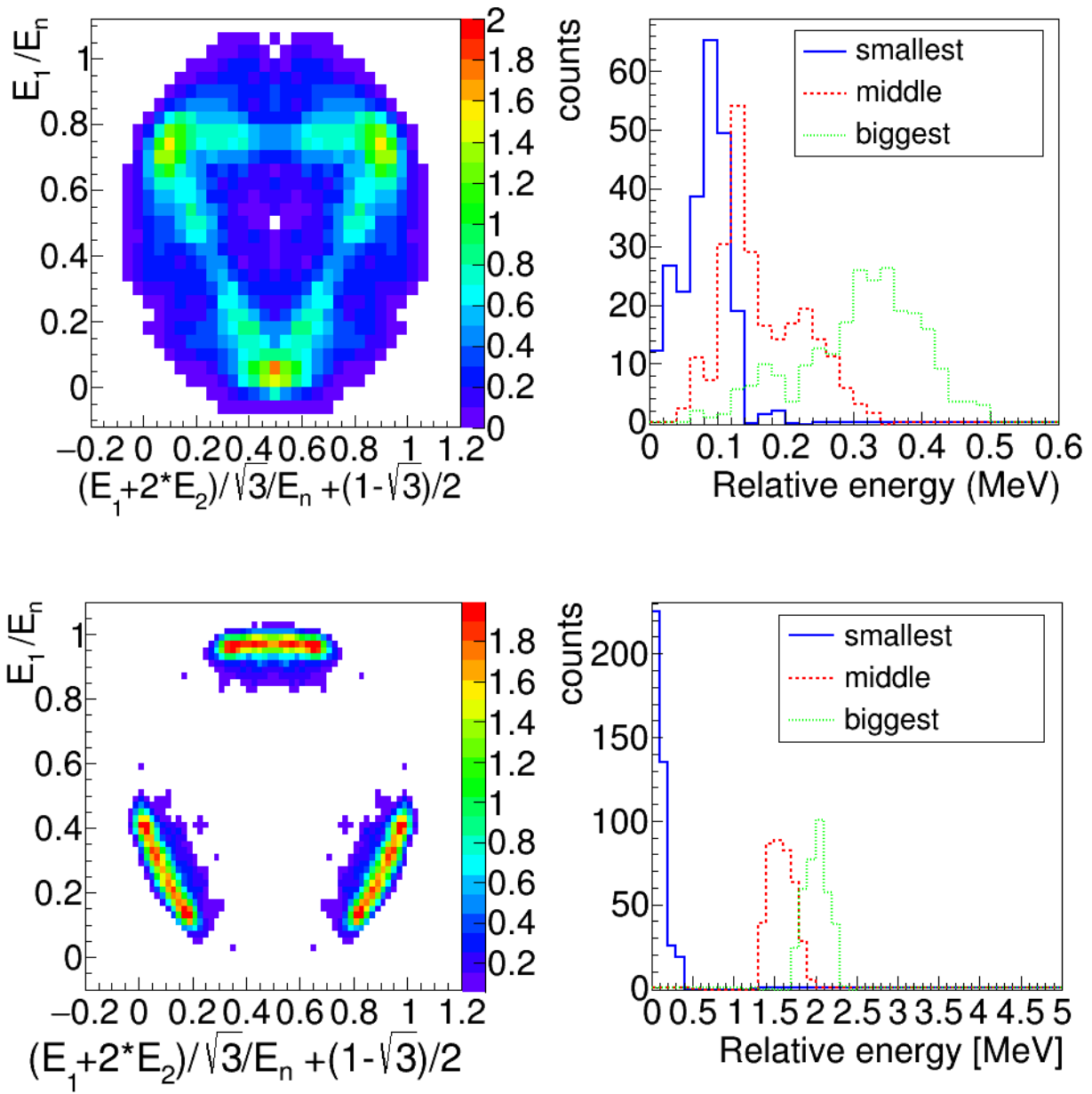

*Figure 11: Experimental data. Left panels, Dalitz plots for the Hoyle state (top) and the (3⁻) state (bottom) after background subtraction. Right panels, relative energies of the three possible couples of alpha particles (Blue solid line, smallest value; green dotted line, biggest value; red dash-dotted line, middle value).*

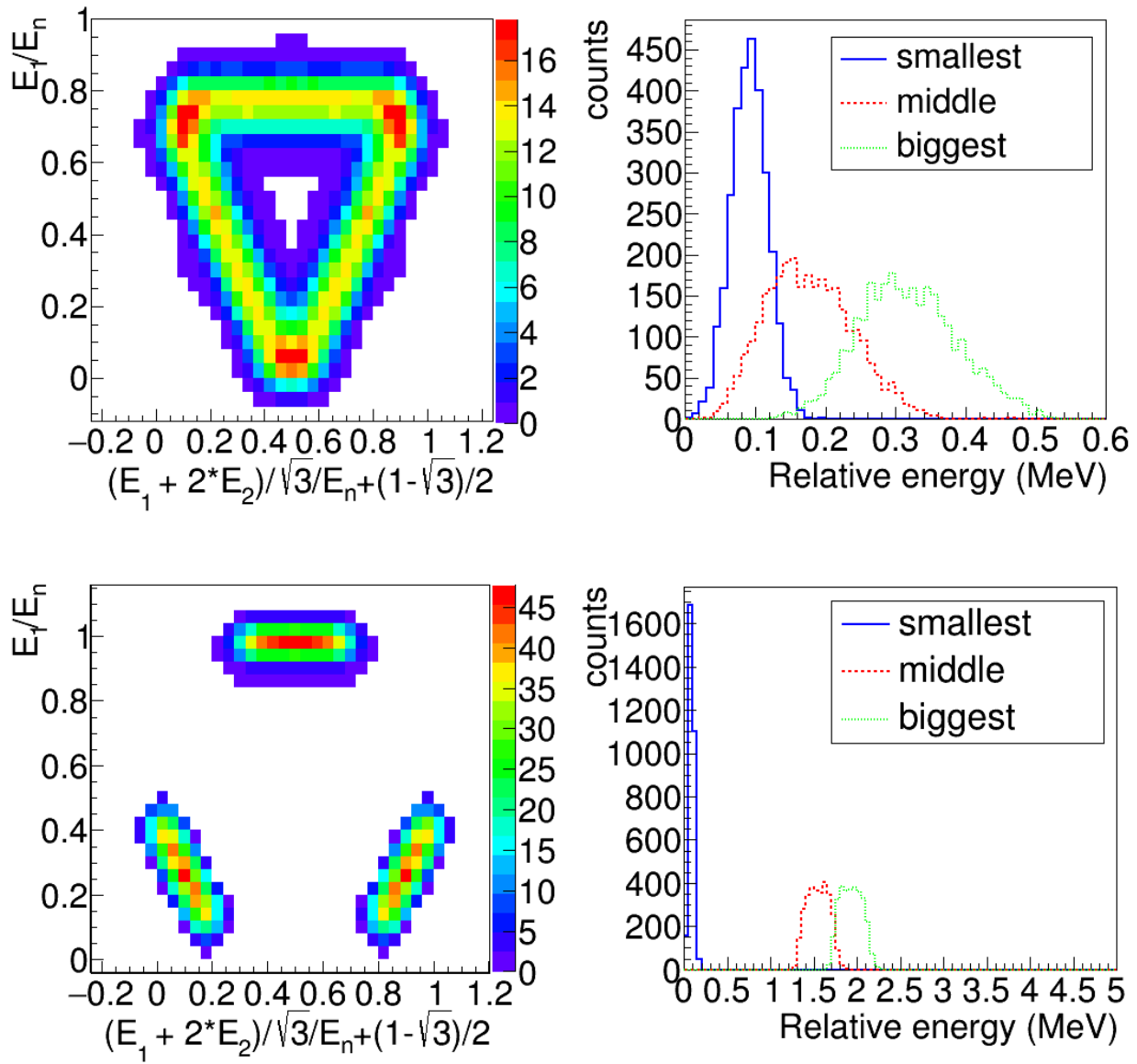

*Figure 12: Monte Carlo simulation of the Hoyle state and the (3⁻) state sequentially decaying through the ground state of $^8$Be. The energy resolution of the relative energy is 50 keV. Panels same as for Figure 11.*

**Alpha multiplicity four events:**
Only 659 events with alpha multiplicity 4 were detected during this experiment. In the analysis we assume these events are coming from the disintegration of the system into 6 alpha particles proceeding through a decay into $^8$Be (in the ground state) and $^{16}$O (with enough excitation energy to decay into 4 alpha particles). The position of the interaction point inside the gas volume is determined with a recursive procedure based on the reaction kinematics, energy and momentum conservation. In each iteration, the kinetic energy of $^{16}$O is obtained from the kinetic energy of the four detected alpha particles, while its excitation energy is obtained from the sum of the kinetic energies of the four alpha particles in their center of mass. In order to avoid the mixing of alpha particles coming from the $^8$Be with those from the $^{16}$O, a Monte-Carlo simulation was performed to determine the energy threshold required to remove the alpha particles from the $^8$Be. Only events with four alpha particles with energy larger than 13 MeV were considered for further analysis. Neither HIPSE nor PACE4 produced alpha multiplicity four events. The background estimated by mixing alpha particles from different events with multiplicity 4 was subtracted from the experimental distribution.

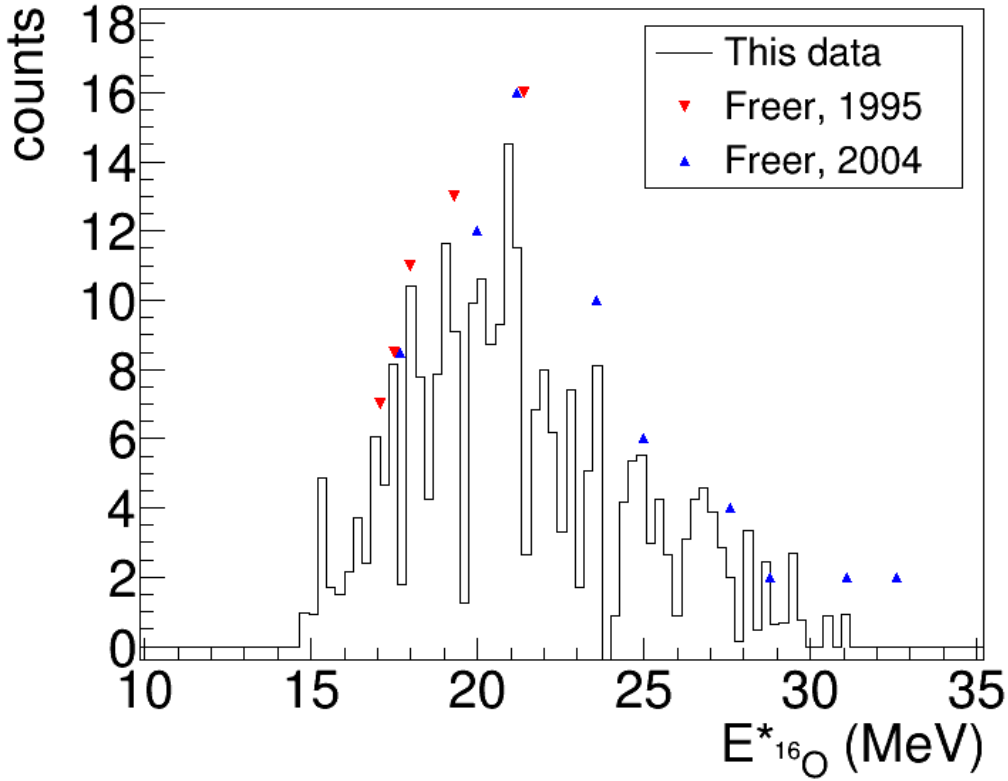

*Figure 13: Reconstructed excitation energy of $^{16}$O obtained from the events with alpha multiplicity 4 after subtracting the background estimated by mixing alpha particles from different events with multiplicity 4. States measured by Freer et al. refs. [13, 30] are also reported.*

The excitation function of $^{16}$O is shown in Figure 13. Neither HIPSE nor PACE4 produced alpha multiplicity four events. The background estimated by mixing alpha particles from different events with multiplicity 4 was subtracted from the experimental data. Funaki et al. [12, 31] predicted a state in $^{16}$O at 15.1 MeV with the structure of the "Hoyle" state in $^{12}$C coupled to an alpha particle. Our excitation function shows 8 events at 15.2$\pm$ 0.2 MeV that could correspond to the $0_6^+$ state in $^{16}$O analogous to the $^{12}$C Hoyle state. This state has been observed in the past by several authors [32-35] and recently by K.C.W Li et al. [36]. The other structures in the excitation function correspond to $^{16}$O states already observed by other authors. Even though the statistics are low, further analysis was performed for the 15.2 MeV group as well as for other peaks. For each peak we determined the amount of events decaying into one alpha particle and one $^{12}$C in the Hoyle state or two $^8$Be in the ground state. To do this we considered the six possible combinations of two alpha particles (0-1, 2-3; 0-2, 1-3; 0-3, 1-2), the decay proceeded through two $^8$Be in the ground state if the relative energies of 0-1 and 2-3 or 0-2 and 1-3 or 0-3 and 1-2 were less than 250 keV. In the same way we considered the four possible combinations of three alpha particles 0-1-2, 0-1-3, 0-2-3, 1-2-3 and checked if the decay proceeded through the $^{12}$C Hoyle state. In this case the sum of the kinetic energies of the three alpha particles in their center of mass should be less than 500 keV or 600 keV for the events in the higher energy states. The relative partial decay widths R were calculated as:

$$R = \frac{\Gamma(^8Be + {}^8Be)}{\Gamma(\alpha + {}^{12}C(0_2^+))} = \frac{Yield(^8Be + {}^8Be)}{Yeld(\alpha + {}^{12}C(0_2^+))}$$

Monte Carlo simulations showed that the detection efficiency for the two decay modes is the same to within a few percent. The results are shown in Table 1 and compared with data from Freer et al. [13]. The ratios R for the states at 17, 19 and 21 MeV agree quite well with those measured by Freer et al. The peak at 15.2 MeV shows the same decay probability into two $^8$Be ground states or α + $^{12}$C Hoyle state. This might indicate that the state has the same characteristics of the $^8$Be ground state and $^{12}$C Hoyle state. In order to increase the statistics we have also re-analyzed the alpha multiplicity four data from our previous experiment performed at 9.7 MeV/nucleon, these data had higher statistics, but were collected with an experimental setup with lower granularity using standard electronics [22]. The excitation energy distribution of $^{16}$O is shown in Figure 14. Due to the lower granularity of the detectors the accuracy on the energies in this excitation function is worse than that of Figure 13, but the same structures can be identified. Figure 15 shows the sum of the energies of the four alpha particles as a function of the reconstructed angle of the $^{16}$O in the laboratory frame, for the data in the 15.2 MeV peak. The data with total kinetic energy less than 120 MeV were selected for further analysis. The events with total kinetic energy larger than 120 MeV were discarded since the reconstructed excitation energy of $^{24}$Mg was very close to or above the maximum available energy. The relative partial widths were calculated as described above and the results are shown in the central column of Table I. Also in this case the ratios for the states at 17, 19 and 21 MeV agree with ref. [13] and the peak at 15.2 MeV shows the same decay probability for the disintegration into two $^8$Be in the ground state and α + $^{12}$C Hoyle state.

For those events the reconstructed excitation energy of $^{24}$Mg peaks at about 34 MeV as shown in Figure 16. It is interesting to note that this value is very close to the energy of 33.42 MeV predicted by Yamada et al. [4] for a state analogous to the $^{12}$C Hoyle state in $^{24}$Mg.

*Table 1: Relative partial decay widths*

| Energy | Γ(Be)/ Γ(Hoyle) this work | Γ(Be)/ Γ(Hoyle) Our previous work [22] | Γ(Be)/ Γ(Hoyle) Freer et al. [13] |
|---|---|---|---|
| 15.2±0.2 | 1±0.7 | 0.96 ± 0.3 | |
| 17.1 | 0.6±0.3 | 0.7 ± 0.3 | 0.65 ± 0.16 |
| 17.5 | | 0.6 ± 0.3 | 0.72± 0.18 |
| 19.7 | 0.43±0.2 | 0.6 ± 0.5 | 0.47± 0.15 |
| 21.4 | 5.3±2.8 | 3 ± 1 | >3± 1.1 |

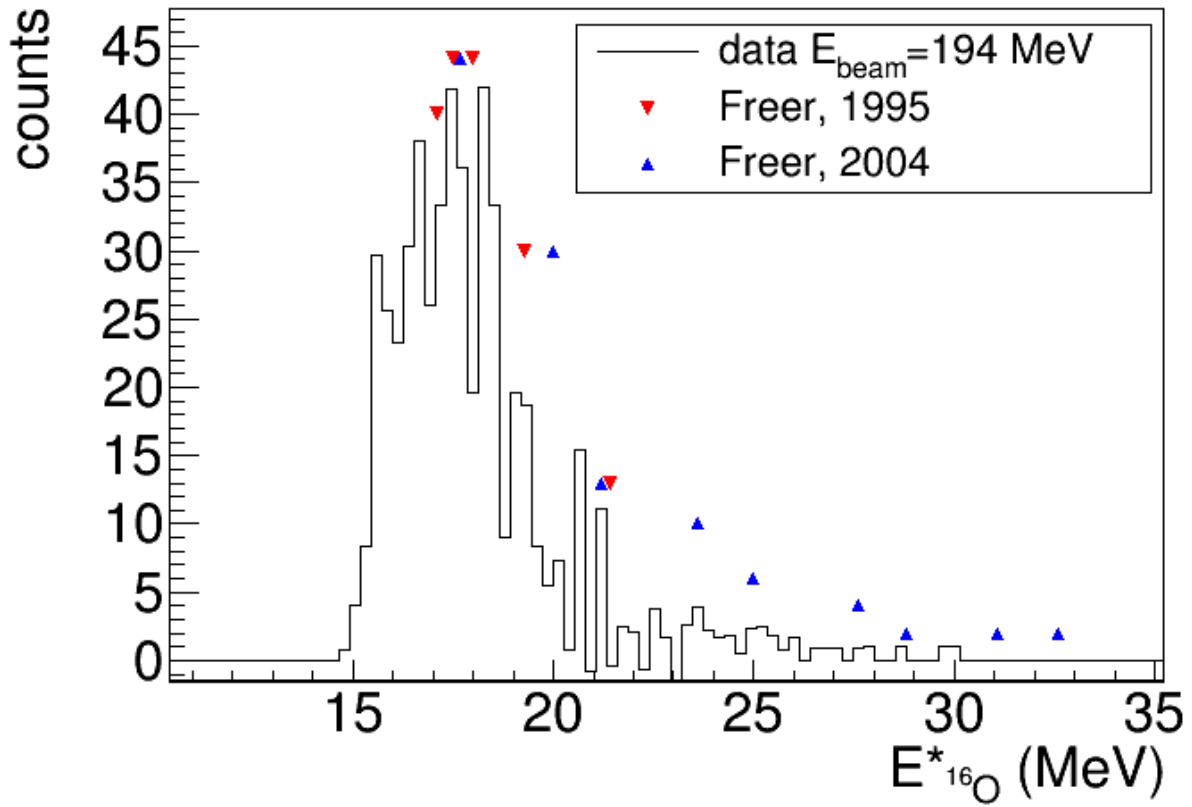

Figure 14: *Reconstructed excitation energy of $^{16}O$ obtained from the events with multiplicity 4 in the run with maximum beam energy 9.7MeV/nucleon. States measured by Freer et al. refs. [13, 30] are also reported.*

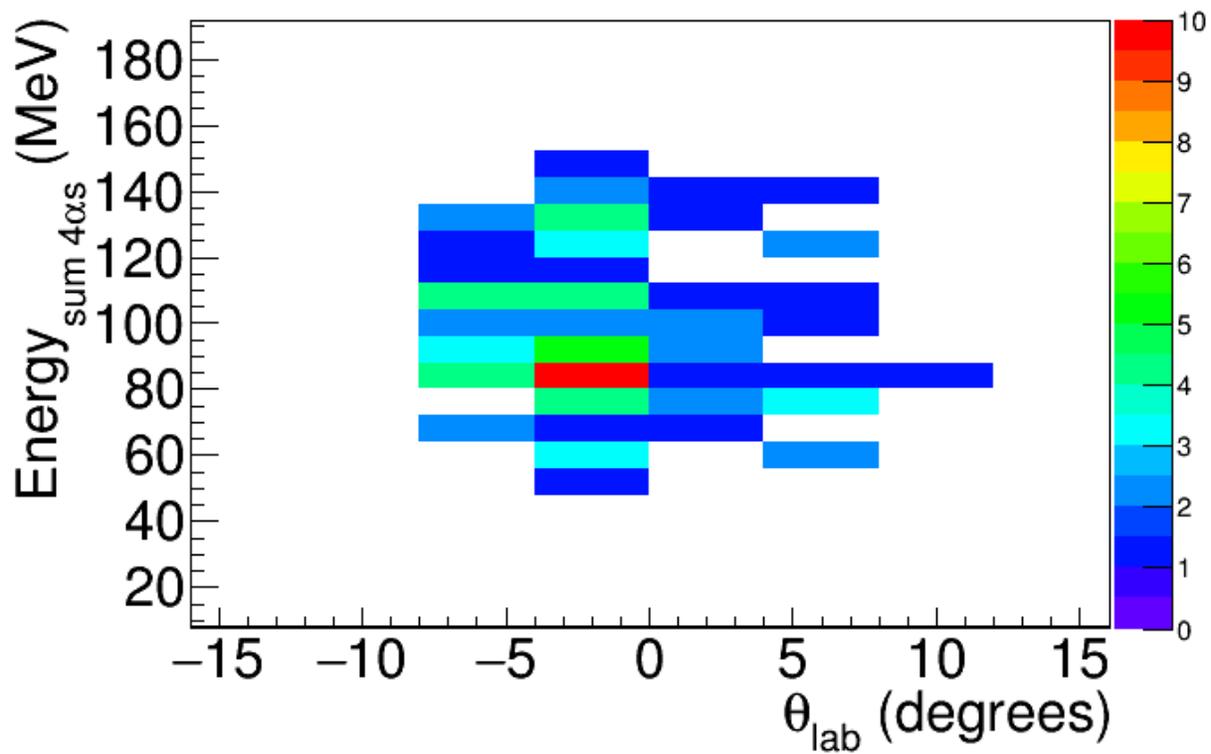

*Figure 15: Sum of the energies of the four detected alpha particles as a function of the reconstructed scattering angle of $^{16}O$ in the laboratory frame.*

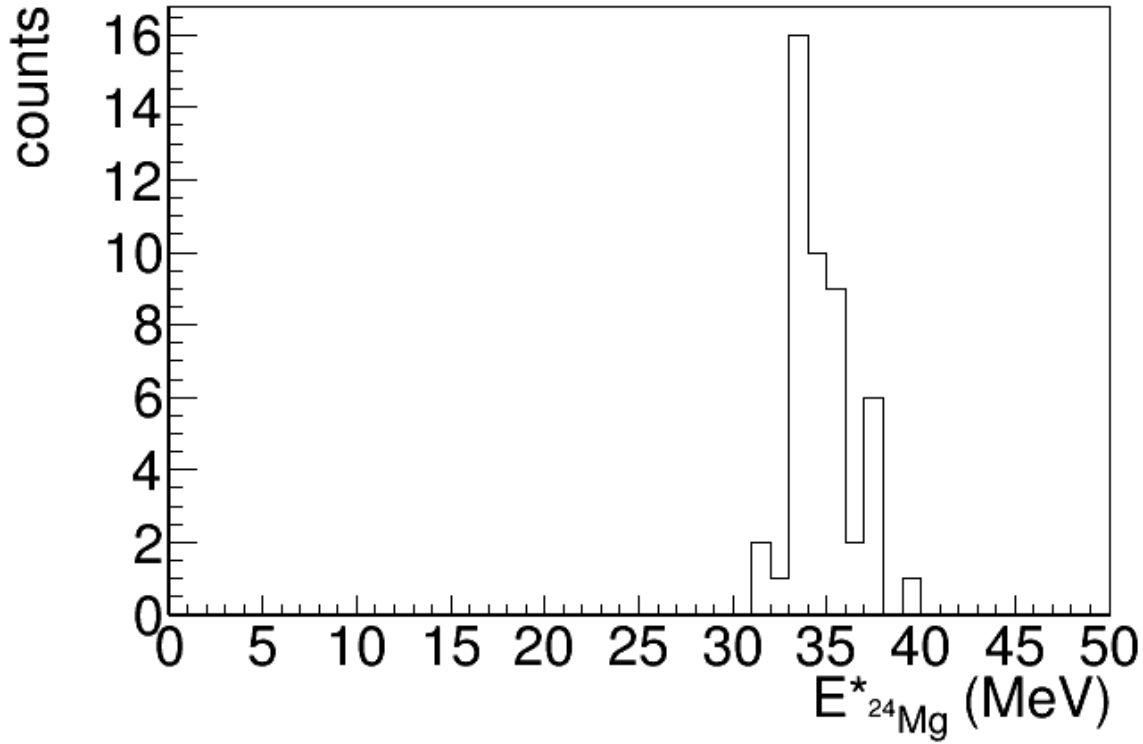

*Figure 16: Reconstructed excitation energy of $^{24}$Mg for the events in the 15.2 MeV peak with total kinetic energy less than 120 MeV.*

**Conclusions:**

In this paper we showed that it is possible to use thick target inverse kinematics to study excited states of light nuclei disintegrating into multiple alpha particles. Excited states of $^8$Be and $^{12}$C were observed, in agreement with previous experimental data. The excitation function of $^{16}$O was also investigated. An interesting structure appears at excitation energy of 15.2 MeV. Although the statistical uncertainty is large, the partial decay width analysis indicates that the events in this state decay with the same probability into two $^8$Be ground states or to an alpha and a $^{12}$C in the Hoyle state. This suggests that this $^{16}$O state can be identified as analogous to the $^{12}$C Hoyle state. Moreover, for those events, the reconstructed excitation energy of $^{24}$Mg peaks at around 34 MeV. Yamada et al. predicted an excited state in $^{24}$Mg analogous to the Hoyle state at about 4.94 MeV above the 6 alpha particles threshold or 33.42 MeV.

In the future we plan to use this same experimental method to collect larger statistics data on $^{16}$O and later to explore heavier systems. We are also considering the possibility to insert an active TPC volume at the end of the chamber in the region after the point where the beam is stopped and before the silicon detector array. This will help to reduce the energy threshold and to trace the trajectories back to the interaction point.

**Acknowledgments**

This work was supported by the United States Department of Energy under Grant # DE-FG03- 93ER40773 and by The Robert A. Welch Foundation under Grant # A0330.

[1] K. Ikeda, N. Takigawa, and H. Horiuchi, 1968, *Prog.Theor, Phys*. Suppl. Extra Number, 464

[2] A. Tohsaki et al. 2001, *Phys. Rev. Lett.*87, 192501

[3] N. Itagaki et al. 2007, *Phys. Rev. C* 75, 037303

[4] T. Yamada and P. Schuck 2004, *Phys. Rev. C* 69, 024309

[5] Tz. Kokalova *et al.* 2006,*Phys. Rev. Lett.*96,192502

[6] M. Itoh *et al.* 2004 *Nucl. Phys. A* 738 268

[7] M. Freer *et al.* 2011 *Phys. Rev. C* 83 034314

[8] M. Itoh *et al.* 2011 *Phys. Rev. C* 84 054308

[9] W. R. Zimmerman *et al.* 2014 *Phys. Rev. Lett.* 110 152502

[10] O. S. Kirsebom et al. 2010, *Phys. Rev. C*, 81, 064313

[11] D. J. Marin-Lambarri *et al.* 2014 *Phys. Rev. Lett.* 113 012502

[12] P. Chevallier *et al.* 1967 *Phys. Rev.* 160, 827

[12] Y. Funaki et al. 2008 *Phys. Rev. Lett.* 101 082502

[13] M. Feer et al. *Phys. Rev. C* 51, 1682

[14] E.G. Adelberger et al. 1970, *Nucl. Phys. A* 143, 97

[15] N. Curtis *et al.* 2013 *Phys. Rev. C* 88, 064309

[16] Tz. Kokalova. *et al.* 2013 *Phys. Rev. C* 87 057307

[17] T. Kawabata et al. Journal of Physics: Conference series 436 (2013) 012009

[18] M. Freer et al. 1998 *Phys. Rev. C* 57, 1277

[19] K. Artemov et al., 1990 *Sov. J. Nucl. Phys.* 52, 406

[20] O.B. Tarasov, D. Bazin, 2008, *Nucl. Instr. and Meth B*, 266, 4657-4664.

[21] D. Lacroix, A. Van Lauwe, D. Durand, 2004, *Phys. Rev. C* 69, 054604

[22] M. Barbui et al.2014 *EPJ Web of Conferences* 66, 03005

[23] SRIM, SRIM The Stopping and Range of Ions in Matter, James F. Ziegler, Jochen P. Biersack, Matthias D. Ziegler, Lulu Press Co. Morrisville, NC, 27560 USA

[24] G.V. Rogachev et al. AIP Conf. Proc., 2010, 1213, 137

[25] R. Abegg and C.A. Davis, 1991, *Phys. Rev C* 43, 2523

[26] M. Barbui et al. 2016, *EPJ Web of Conferences* 117, 07013

[27] R.Smith et al. 2017, *Phys. Rev. Lett*. 119, 132502

[28] D. Dell'Aquila et al. 2017, *Phys. Rev. Lett.* 119, 132501

[29] Tz. Kokalova et al., 2013, *Phys. Rev. C* 87, 057307

[30] M. Freer et al. 2004 *Phys.Rev. C,* 70, 064311

[31] Y. Funaki et al. 2010, J. Phys. G: Nucl Part. Phys. 37, 064012

[32] T.Marvin and P.Singh, 1972 *Nucl. Phys.A* 180, 282

[33] A. Fraweley, K. Bray and T. Ophel, 1978 *Nucl. Phys. A* 194,161

[34] O.Bilaniuk, H. Foutune and R. Middleton, 1978*Nucl. Phys. A* 305, 63

[35] L.L. Ames 1982 *Phys. Rev. C* 25, 729

[36] K.C.W. Li et al, 2017 *Phys. Rev. C* 95, 031303